**Extreme timescale core-level spectroscopy with tailored XUV pulses**

R. Singla[1], D.C. Haynes[1], K. Hanff[2], I. Grguraš [1,3], S. Schulz[1,4], H.Y. Liu[1], A. Simoncig[1,5], F. Tellkamp[1], S. Bajt[4], K. Rossnagel[2], A.L. Cavalieri[1*]

[1] Max-Planck Institute for the Structure and Dynamics of Matter, Luruper Chaussee 149, 22761 Hamburg, Germany.

[2] Institute of Experimental and Applied Physics, Kiel University, Olshausenstraße 40, 24098 Kiel, Germany

[3] Class 5 Photonics GmbH, Notkestraße 85, 22607 Hamburg, Germany

[4] Deutsches Elektronen-Synchrotron DESY, Notkestraße 85, 22607 Hamburg, Germany.

[5] Elettra Sincrotrone Trieste SCpA, Str Statale 14 Km 163-5, I-34149 Trieste, Italy

[*] *corresponding author*

**Abstract**

A new approach for few-femtosecond time-resolved photoelectron spectroscopy in condensed matter that balances the combined needs for both temporal and energy resolution is demonstrated. Here, the method is designed to investigate a prototypical Mott insulator, tantalum disulphide (1$T$-TaS$_2$), which transforms from its charge-density-wave ordered Mott insulating state to a conducting state in a matter of femtoseconds. The signature to be observed through the phase transition is a charge-density-wave induced splitting of the Ta 4f core-levels, which can be resolved with sub-eV spectral resolution. Combining this spectral resolution with few-femtosecond time resolution enables the collapse of the charge ordered Mott state to be clocked. Precise knowledge of the sub-20-femtosecond dynamics will provide new insight into the physical mechanism behind the collapse and may reveal Mott physics on the timescale of electronic hopping.

**Introduction**

Ultrafast optical spectroscopy can be used to investigate and control microscopic dynamics in complex materials. Few-femtosecond (fs) temporal resolution is essential to disentangle intertwined degrees of freedom in correlated systems.[1,2] In a complementary approach time-resolved core-level photoelectron spectroscopy (PES) can be used to interrogate element-specific dynamics. Using PES, the electronic features of atoms directly involved in certain emergent phenomena can be isolated and observed, provided that the bandwidth of the probe used to eject photoelectrons is narrow enough to resolve the relevant electronic features. In conjunction with the time-energy uncertainty principle, this imposes constraints on the maximum time resolution that can be achieved.

Here, an approach for ultrafast PES that provides the flexibility to tune the bandwidth and corresponding time resolution is reported. With an optimised probe, electronic dynamics in correlated systems can be accessed with the highest possible temporal resolution. By determining the relevant timescales, underlying driving mechanisms behind important phase transitions in materials can be identified and isolated. For example, this tool could be applied in graphene: by observing the initial few-fs dynamics of its non-thermal carrier distribution, the carrier multiplication process could be unveiled.[3,4]

Dynamics in the transition metal dichalcogenide tantalum disulphide (1$T$-TaS$_2$), which exhibits a Mott insulating state at low temperatures, can also be investigated using this approach. In 1$T$-TaS$_2$ a charge density wave (CDW) structure coincides with the Mott insulating state. The CDW state can be melted with a short optical laser pulse, leading to a phase transition in the material from an insulating to conducting state.[5] A direct signature of the CDW is splitting of the Ta 4f core-levels; time resolved PES can be employed to observe this splitting as a function of time.

In order to perform this experiment, the energy resolution of the extreme ultraviolet (XUV) probe must be sufficient to resolve the core-level split which is less than one eV. Additionally, the time resolution should be higher than the fastest vibrational mode in order to disentangle electronic from structural dynamics. Past experiments on 1$T$-TaS$_2$ at free-electron laser sources[6] or table-top XUV sources[7,8] have had insufficient time resolution to investigate these dynamics. Conversely, attosecond sources with sufficient time resolution do not have the energy resolution required to identify the spectroscopic signatures of the CDW state. Alternatively, attosecond sources that would have sufficient time-resolution do not have the energy resolution required to identify the spectroscopic signatures of the CDW state. Typical attosecond sources have several eV of bandwidth which would obscure the core-level splitting. The new approach described below is unique in that it affords sufficient resolution in both temporal and spectroscopic domains to clock the collapse of the CDW-ordered Mott state in 1$T$-TaS$_2$ on the fundamental timescale of electron hopping. This timescale will provide new insight into the precise mechanism underlying the phase transition.

**Optical Drive Laser**

In addition to an appropriate XUV probe, an optical pump pulse is also required to initiate dynamics in correlated systems. To preserve the time resolution provided by the probe pulse, the pump must be similar in duration. The CDW core-level splitting in 1$T$-TaS$_2$ has a magnitude of approximately 700 meV, and the time resolution in the XUV probe is therefore constrained to a maximum of approximately 2 fs by the time-energy uncertainty principle. As a result, a near single-cycle optical laser pump pulse is required. In practice a single laser system is used to generate both the few-femtosecond single cycle pump laser pulse as well as the 1 eV XUV probe pulse. High-harmonic generation (HHG)[9,10] is used to up convert the optical laser light to the XUV regime.

At the core of the laser system is a 3-kHz repetition rate Ti:sapphire chirped pulse amplification (CPA) amplifier that is seeded with waveform stabilised, or carrier-envelope phase stabilised (CEP) oscillator laser pulses.[11] CEP refers to the varying phase of the electric field with respect to the pulse amplitude envelope and the CEP frequency, $v_{CEP}$, represents the proportional change in the phase between consecutive pulses. For example, a CEP frequency equal to half of the pulse repetition rate corresponds to a CEP shift of π between each pulse.

The CEP frequency can be determined using optical frequency comb measurements in a self-referenced f-to-2f scheme. The lower-frequency part of the spectrum is measured, before being frequency-doubled and compared with the high-frequency part of the original comb. This allows the beat note between the frequency components in the fundamental and second harmonic to be measured, which is the CEP frequency. The method requires an octave-spanning spectrum so that the frequency-doubled part of the spectrum can be compared with the high-frequency part, and several schemes exist for extending the spectrum to meet this requirement.[12] Waveform stabilisation is crucial because the highly nonlinear process of high-harmonic generation is field dependent. In general, for very short driving optical laser pulses it can be expected that the XUV emission will vary substantially as a function of the driving waveform.

To make the shortest possible drive laser pulses, which are fundamentally limited by the carrier wavelength of the laser, a hybrid compression scheme is employed.[13] In this scheme, pulse compression following amplification in the CPA amplifier is achieved in distinct steps. The majority of the compression is performed in a prism compressor with approximately 2 m distance from apex to apex, while the final compression is performed in a positive dispersion chirped mirror compressor. Chirped mirrors introduce dispersion by reflecting different wavelengths from different depths within the dielectric mirror coating.[14] The chirped mirror compressor is required because it allows for compression without propagation in bulk material, mitigating the effects of self-phase modulation (SPM) that would otherwise have led to spectral narrowing of the amplified laser pulse. Using this scheme, 1mJ pulses are delivered with 72 nm bandwidth FWHM and a pulse duration of 22 fs FWHM.

To generate even shorter pulses more laser bandwidth is required. Therefore, after the hybrid compressor, the ~1 mJ, 22 fs laser pulses are focused into a hollow core fibre (HCF) filled with neon gas at an absolute pressure of 1.6 bar. The core diameter of the fibre is 250 µm and the length is 105 cm. The HCF acts as a waveguide to keep the intensity of the laser pulse nearly constant as it interacts with the neon gas. The interaction of the relatively intense laser pulses with the background gas leads to SPM. In this case, however, SPM leads to spectral broadening rather than narrowing as would have been the case in the prism compressor without the hybrid scheme. The difference here is that the input pulse is either transform limited or slightly positively chirped (red frequency components lead blue frequency components), in contrast to the pulses in the prism compressor of the amplifier which are negatively chirped when propagating in the bulk material.

SPM is a result of the optical Kerr effect, and occurs in media whose refractive index varies with incident pulse intensity. The phase of the pulse is shifted and its frequency spectrum broadened. In addition to SPM, the pulse is also subject to linear dispersion while propagating in the HCF, which stretches the pulse in time. To compensate for the dispersion and compress the pulses to the few-femtosecond level, a negative dispersion chirped mirror compressor is used.[15] Following final compression, nearly transform limited pulses with a duration of 3.3 fs FWHM are delivered with pulse energies up to 600 µJ. A transient-grating frequency resolved optical gating measurement device (TG-FROG)[16] was used to characterise the optical pulse. The TG-FROG is intrinsically phase-matched for all optical frequencies and can therefore accommodate the HCF broadened spectrum that spans more than one octave.

As previously mentioned, the HHG process is sensitive to the waveform of the compressed pulse. The dependence becomes more pronounced as the pulse gets closer to the single-cycle limit. Although the amplifier is seeded with CEP-stable oscillator pulses, phase drifts are accumulated in the amplification process. Therefore, the CEP of the amplified pulses[17] is measured again following spectral broadening in the hollow core fibre. In this case, measurement of the CEP phase is simplified due to the fact that the broadened spectrum is already octave spanning. After the CEP is evaluated, feedback is applied on the stretcher in the CPA amplifier. Using this scheme, the CEP is typically stabilised to a level of 60 mrad. Additional feedback loops have been implemented to allow this level of short-term phase stability to be maintained for several days, a time period significantly longer than the measurement cycle.

The long-term stability of the laser system, including the pulse energy and compression of the few-femtosecond laser pulse is extremely sensitive to the temperature of the laser environment. Small changes in cooling water temperature, for example, couple into the CPA pump laser output power. If these changes are systematic the effects can be highly detrimental to sensitive measurements and could in potentially present as a periodic artefact that is easily confused with a real signal. Therefore, significant effort has been made to actively stabilise both the air temperature and cooling water temperature.

The air temperature is controlled to within 0.5 K using a commercial flow-box system. The cooling water temperature for the CPA pump laser, oscillator pump laser, as well as the laser breadboard and optics are stabilised to within 5 mK using a custom design. Such precise control is facilitated by the addition of a fast heating element to the cooling water circuit. The fast heater allows temperature drifts and fluctuations from the water chiller to be smoothed – provided that the final temperature is slightly higher than the set point of the water chiller. In addition to stabilising the temperature to within 5 mK, the temperature set point can be chosen to compensate for changing thermal loads or environmental drifts. For example, if a new heat source is introduced on the optical table, the cooling water temperature set point is automatically reduced to keep the end temperature of the optical table constant. The advantages of the temperature stabilisation are two-fold. With proper choice of computer control parameters, both fast fluctuations and systematic or periodic drifts can be eliminated.

**Frequency Conversion by HHG**

Three-femtosecond, waveform stabilised optical pulses are suitable for impulsively pumping dynamics in correlated systems. However, to use them to probe core-level dynamics with time resolved PES, they need to be frequency converted to the XUV region via HHG. In a classical picture, when an intense optical laser pulse is incident on a gas target, electrons are ionised from the gas atoms. These electrons are then accelerated away from the core ion by the overlapping laser pulse. When the sign of the field reverses, the electron is accelerated back toward the parent ion. If the electron recombines with the parent ion it releases its excess kinetic energy – gained while accelerating in the laser field – as high energy XUV photons. Due to the periodic nature of the laser field that drives the process and the symmetry of the gas target, odd-harmonics of the driving laser wavelength are emitted.[9]

In practice, the laser beam is focused with a 500 mm focal length concave mirror onto a dilute neon gas target. Laser intensities exceeding $10^{15}$ W/cm$^2$ can be applied due to high contrast in the optical pulse, which minimises ionisation depletion of the HHG target gas.[18] With these intensities, XUV cutoff photon energies exceeding 250 eV can be reached, as expected based on the well-known cutoff law[9,19] for the highest energy emitted photon: $E_{cutoff} = 3.17 U_p + I_p$, where $U_p$ is the ponderomotive potential of the streaking laser pulse. The near Gaussian XUV beam profile resulting from the HHG process is shown in Figure 1a. In Figure 1b, the emitted XUV spectrum is displayed as a function of the CEP, revealing a clear periodic structure. The structure is due to variation in the electric field extrema between consecutive half-cycles of the near single-cycle laser pulse, resulting in significantly different electron trajectories. As the cutoff in the XUV spectrum produced by each half-cycle is dependent on the extremum of its electric field, the $\pi$-periodic repeating structure observed in Figure 1b has been called a "half-cycle cutoff."[20]

The XUV emission spectrum is extremely broadband and smooth in the cutoff region. The lack of harmonic structure is caused by neighbouring half-cycles of the laser field having different amplitudes, while the smooth XUV continuum is due to the fact that only one half-cycle of the driving laser pulse has a large enough amplitude to produce such high energy XUV photons. A larger difference between the amplitudes of neighbouring half-cycles leads to a broader XUV continuum. This continuum is well-suited for attosecond streaking applications. In this case, an XUV filter is used to slice a portion of the smooth XUV spectrum, resulting in an isolated XUV pulse with a duration that depends primarily on the bandwidth of the sliced spectral region.[21]

**Attosecond streaking spectroscopy**

To evaluate the overall stability and performance of the system an attosecond streaking experiment was performed in Ne.[22] For this purpose an XUV multilayer mirror with 3.4 eV FWHM bandwidth centred at 164 eV was used to isolate individual attosecond pulses. Additionally, the HHG focusing geometry, backing pressure in the HHG target and compression and CEP of the driving optical laser pulse were chosen to favour the spectral region of the XUV emission spanning the reflectivity curve of the multilayer mirror. After the attosecond pulses were isolated they were focused onto another neon gas target, emitting photoelectrons in a single photon process. The resultant static photoemission spectrum of the neon gas is shown in Figure 2a. With electron binding energies of 21.6 eV and 21.7 eV for the 2p states and 48.5 eV for the 2s state, the emission peaks are evident at approximately 115 eV and 142 eV kinetic energy. The individual 2p states cannot be resolved due to the bandwidth in the attosecond pulse.

In the classic attosecond streaking experiment, a portion of the laser pulse used for HHG is recycled and used to modulate or streak the kinetic energy of the attosecond XUV photoemission. Both pulses are focused onto a neon gas target and temporally and spatially overlapped. The XUV pulse ionises the gas target, producing a burst of photoelectrons with a temporal profile that replicates the profile of the incident photon pulse. The final kinetic energy of the electrons that comprise the photoelectron burst is subsequently increased or decreased depending on their exact time of release into the waveform stabilised laser field. Provided that the duration of the XUV pulse or photoemission burst is less than the half-cycle of the streaking laser field, the final kinetic energy, $E_\text{f}$, for an electron released at $t_0$ and observed parallel to the laser electric field, in atomic units, is given by:[23]

$$E_\text{f}(t_0) = E_i - p_i A(t_0) + \frac{A^2(t_0)}{2}.$$

Here $E_i$ and $p_i$ are the initial kinetic energy and momentum, respectively, of the electron, and $A(t_0)$ is the laser vector potential at the instant of ionisation. Additionally, for a burst of photoelectrons emitted over a finite period of time, the photoelectron spectrum is broadened depending on the duration of emission and strength of the streaking field. As a result, a spectrogram comprised of streaked spectra recorded over the full range of delays between the attosecond XUV pulse and laser pulse can be used to fully characterise both the laser and attosecond XUV pulse. A characteristic streaking spectrogram is shown in Figure 2b. Using standard retrieval algorithms, the attosecond XUV pulse duration was found to be approximately 500 as and the streaking laser pulse duration approximately 3.5 fs, in good agreement with the optical TG-FROG measurements.

**XUV tuning**

While the attosecond streaking experiment provides an important benchmark, the main goal is to perform optical pump-XUV probe measurements on materials with simultaneous access to high resolution in both the energy and time domains. For this purpose, the photon energy and bandwidth of the XUV probe can be tuned by implementing a variety of different multilayer optics. The concept is illustrated in Figure 3, which shows neon photoemission spectra produced by selecting different 'slices' of the full HHG XUV spectrum.

For materials research, the goal is to limit the bandwidth in the XUV probe pulse to less than 1 eV, resulting in approximately 2-femtosecond XUV pulses that can still be used to spectrally resolve electronic features. Higher energy resolution probes may be desirable, but there are fundamental limitations on the contrast between the material pairs used in the multilayer, as well as limits on the number of layers that can effectively contribute due to absorption. As a result, multilayer optics with reflective bandwidths significantly below 1 eV have not been realised. This makes it challenging to resolve the signature splitting of ~700 meV in the 4f core-levels of 1$T$-TaS$_2$.

To further reduce the bandwidth, the XUV spectrum that is incident on the multilayer must also be manipulated. With a near single-cycle drive laser the XUV emission is smooth from approximately 140 eV to the cutoff energy, as only a single half-cycle of the drive laser pulse can produce these photons. Around 100 eV, however the XUV spectrum presents a weakly modulated harmonic structure, which is the hallmark of a multi-cycle process. The XUV spectrum is affected by three contributing factors: the carrier envelope offset phase of the drive laser pulse, the carrier wavelength of the single-cycle drive laser pulse, and the phase matching conditions in the HHG target. If these parameters are chosen correctly and a suitable multilayer optic is used to slice the XUV spectrum, the resultant bandwidth in the XUV probe can be further reduced when the weakly modulated XUV spectrum is aligned with the reflectivity peak.

The influence of the CEP on the emitted XUV radiation is shown in Figure 4a. The corresponding effect is observed in the bandwidth of the neon 2p photoemission peak, as shown in Figure 4b. Here, there is a clear periodic broadening of the 2p photoemission peak with π-periodicity in the CEP. Depending on the phase, the bandwidth in the XUV probe alternates between ~750meV and ~2 eV. Therefore, to observe the spectroscopic signature of the CDW in 1$T$-TaS$_2$ the correct CEP phase must be identified and stabilised for the duration of a time resolved measurement. This places additional pressure and or constraints on the accuracy and endurance of the CEP stabilisation loops, as well as the long-term stability of the CPA-amplifier and HCF compression systems.

**Photoelectron spectroscopy in 1$T$-TaS$_2$**

The key to clocking the collapse of the CDW-ordered Mott state in 1$T$-TaS2 is observation of the splitting of the Ta 4f core-level states with sufficient temporal and energy resolution. With time resolution higher than the speed at which the crystal lattice can move, electronic and structural dynamics can be unambiguously disentangled. Using the tailored XUV probe described above, sufficient energy resolution can be achieved. A high-resolution reflectivity curve of an appropriate multi-layer mirror obtained using a synchrotron XUV source is shown in Figure 5a. When combined with a tuned XUV spectrum, the Ta 4f-core-level splitting in 1$T$-TaS$_2$ has now been resolved, as shown in Figure 5b. This is an important demonstration, but proof of the spectral resolution is only one piece of the puzzle.

The time resolution that can be achieved with this probe must also be characterised. For this, a streaking experiment was performed. However, as the XUV pulse duration is expected to be approximately 2 fs, a streaking spectrogram like that in Figure 2 is not obtained. Rather than observing a clear shift of the photoelectron kinetic energy, sidebands on the main photoemission peak are expected, due to interference between different half cycles of the laser field overlapping with the temporally extended photoemission. The spacing between the sidebands is given by the photon energy of the pump laser. Under these conditions, a time-resolved spectrogram obtained by scanning the overlap between the optical pump laser pulse and 2-fs XUV probe is shown in Figure 6, panel a. Here sidebands are clearly evident at the overlap between the XUV pulse and peak of the optical pump laser pulse. The overall temporal resolution that was achieved in this experiment – which can correspondingly be expected in a time resolved measurement of the insulator to metal phase transition in $1T$-TaS$_2$ – is evaluated by calculating the width of the broadened 2p photoelectron spectrum as a function of delay. The temporal extent of the broadened sidebands feature shown in Figure 6b is a convolution of the XUV pulse duration with the pump laser pulse duration. Here the overall temporal resolution is approximately 6 fs FWHM. This result is in agreement with expectations based on the laser pulse duration and bandwidth of the XUV probe pulse. Moreover, a resolution of ~6 fs is sufficient to freeze out even the fastest lattice dynamics that are expected to occur on the 20-femtosecond timescale.

**Conclusion**

Observation of dynamics in correlated systems and condensed matter in general requires a delicate balance between spectroscopic and temporal resolution. While there has been a great deal of effort to observe attosecond dynamics in atoms, molecular and even condensed matter systems, these sources are not immediately applicable in many interesting cases due to their inherent broad bandwidth. Such bandwidth obscures spectroscopic features that need to be resolved and tracked. In this work, we present a new approach to time-resolved photoelectron spectroscopy that builds on techniques original developed for attosecond lasers and attosecond spectroscopy. Weakly modulated harmonic structure below the cutoff region in the HHG XUV spectrum is tuned and stabilised to further increase the spectroscopic resolution that can be achieved using muti-layer XUV optics. In this way, the CDW splitting in $1T$-TaS$_2$ has now been resolved. Simultaneously, an overall experimental time resolution of approximately 6 fs has also been achieved.

In the future, combined with stabilisation systems that decouple the experiment from environmental factors, this approach will allow for time-bandwidth limited spectroscopy in $1T$-TaS$_2$ as well as other relevant correlated electron materials. Furthermore, if the system is combined with suitable far-infrared or THz pump sources, it may provide a realistic means to observe and control the physical properties of these materials with light, opening the door to a new generation of light driven devices.

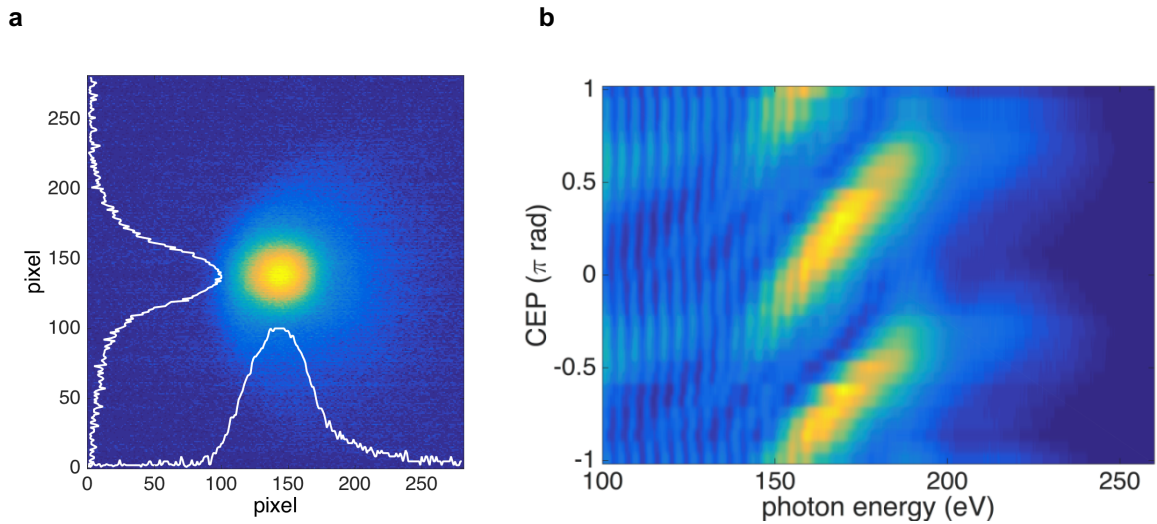

**Figure 1| Broadband high-harmonic XUV emission.** The XUV beam is generated by high-harmonic generation (HHG) in a neon gas target. Panel **a** shows the diffraction-limited beam profile measured at a distance of 4 metres. Panel **b** shows the XUV spectrum as a function of the carrier-envelope phase (CEP) of the near-single-cycle optical laser pulse used to drive the HHG process. Strong variation in the XUV spectrum with periodicity of $\pi$ in the CEP, a feature also known as a "half-cycle cutoff", is clearly visible and confirmation of a very short optical drive laser pulse.

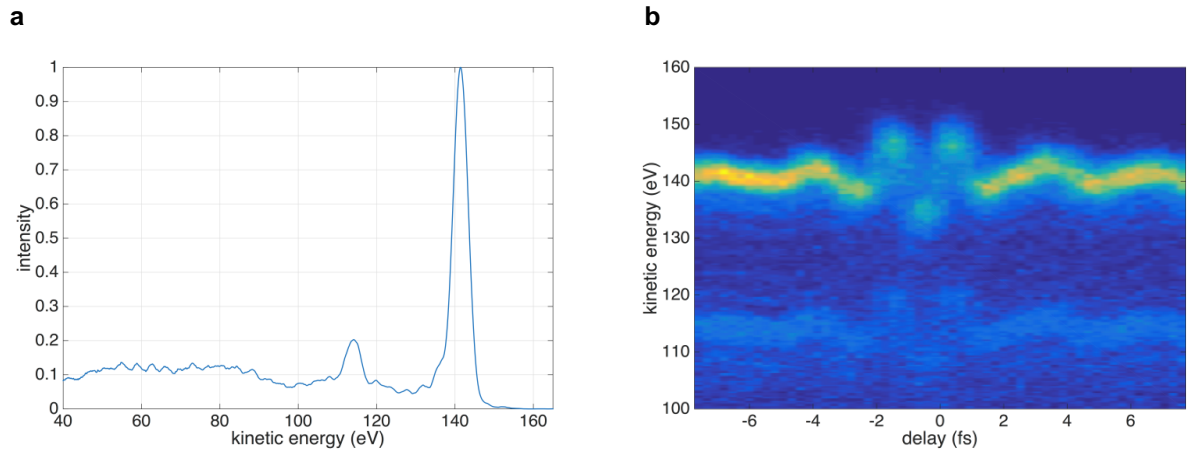

**Figure 2| Attosecond streaking spectroscopy.** A multilayer mirror with reflectivity curve centred at 164 eV in photon energy and with 3.4 eV FWHM bandwidth is used to slice an isolated attosecond XUV pulse. Panel **a** shows the static photoelectron spectrum (PES) obtained from neon gas. Both the 2s and 2p peaks are clearly visible at ~115 eV and ~140 eV kinetic energy, corresponding to binding energies of ~21.7 eV and 48.5 eV respectively. An attosecond streaking spectrogram, recorded as the delay between attosecond XUV pulse and streaking laser pulse is scanned, is shown in panel **b.** Streaking is evident in both the 2p (upper) and 2s (lower) peaks. Analysis of the spectrogram indicates that the attosecond pulse is ~500 as FWHM in duration and that the streaking laser pulse is 3.3 fs FWHM.

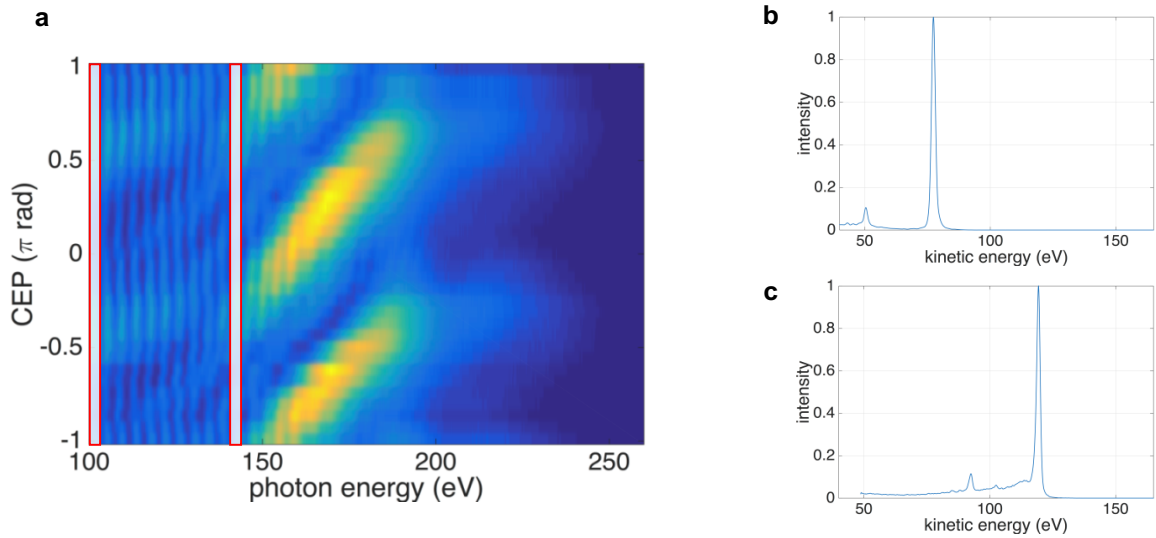

**Figure 3| Tailoring the probe bandwidth and central energy with multilayer optics.** Panel **a** shows the full broadband XUV spectrum as a function of CEP. The two shaded regions correspond the reflectivity curves of different multilayer optics. XUV pulses isolated using these optics have distinct characteristics. The choice of multilayer optic and therefore XUV probe pulse depends on the experimental application. The XUV pulse used to generate the photoemission curve in panel **b** has a central energy of 99 eV and a FWHM of ~1 eV, while that used in panel **c** has a central energy of 141 eV and a FWHM of ~1.1 eV.

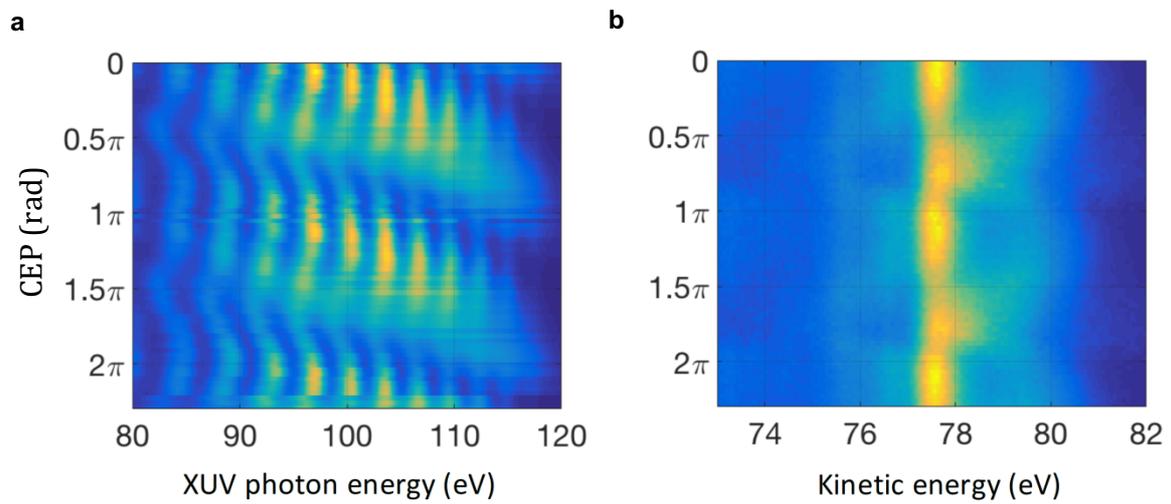

**Figure 4| Carrier-envelope phase dependence of the XUV probe.** Panel **a** shows the CEP-dependence of the XUV emission in the spectral region of interest around 100 eV. In panel **b** shows Ne 2p photoemission spectra as a function of CEP. The Ne 2p photoemission peak bandwidth can be used as a direct measure of the XUV probe bandwidth. It alternates between narrow and broadband pulses with a CEP-period of π. The minimum bandwidth observed is approximately 1 eV FHWM. The maximum bandwidth observed is greater than 2 eV FWHM.

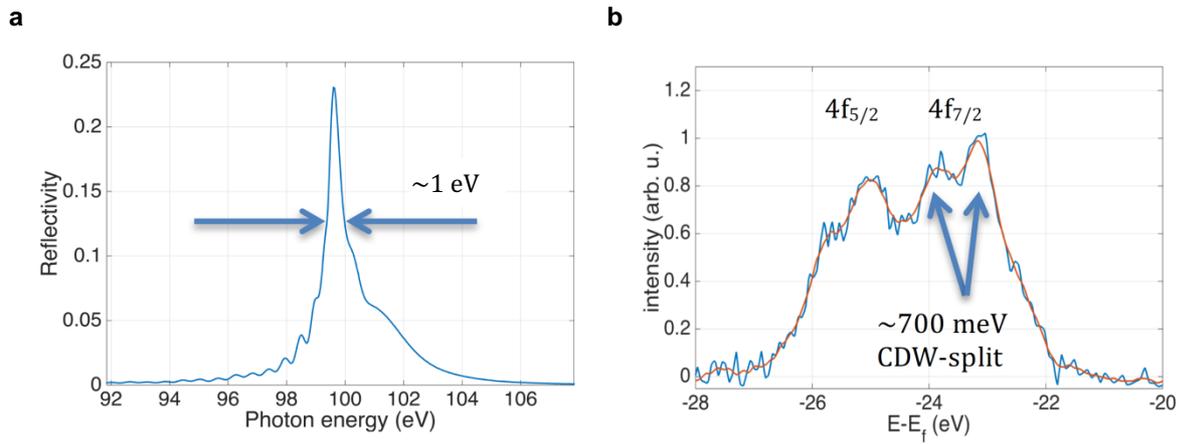

**Figure 5| 1T-TaS$_2$ charge density wave resolved.** The reflectivity curve of a narrowband ~1 eV molybdenum/silicon mirror is shown in panel **a**. The mirror was used to slice XUV pulses from the broadband XUV emission in the weakly structured region of the spectrum below the smooth cutoff region. Tuning the XUV spectrum to align a harmonic peak with the mirror reflectivity optimised the energy resolution of the XUV probe pulse. These pulses were used to probe 1T-TaS$_2$, resulting in the spectrum shown in panel **b**. Characteristic splitting of the Ta 4f peaks is indicative of the insulating charge-density-wave state.

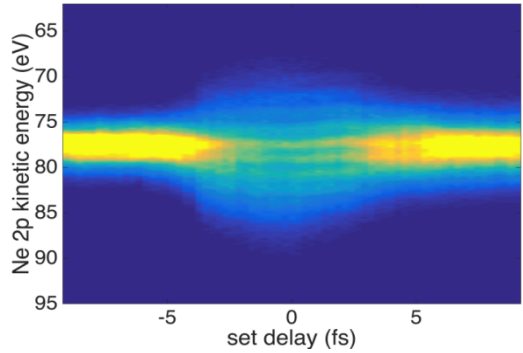 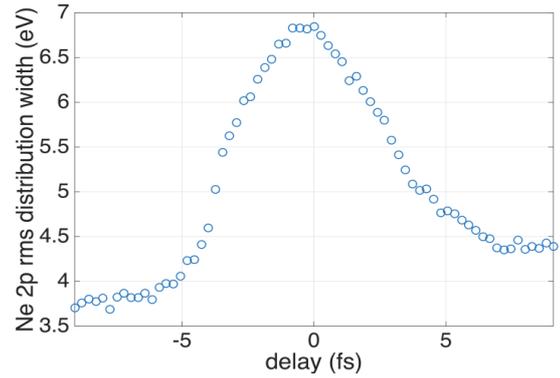

**Figure 6| Pump-probe time resolution.** Cross-correlation between narrowband 99 eV XUV pulses and 3.3 fs optical pump pulse. Panel **a** shows a spectrogram of the neon 2p peak recorded as the delay between the two pulses was smoothly varied. In the region of overlap sidebands appear with spacing given by the pump laser photon energy. Panel **b** shows the statistical width of the 2p peak as a function of delay. The temporal extent of the sidebands feature is ~6 fs FWHM giving a measure of the overall time resolution that can be expected in the experiment.